\documentclass[%
 reprint,
superscriptaddress,
nofootinbib,
 amsmath,amssymb,
 aps,
prd,
]{revtex4-1}

\usepackage{graphicx}
\usepackage{dcolumn}
\usepackage{bm}
\usepackage{hyperref}
\usepackage{float}

\begin{document}

\preprint{APS/123-QED}

\title{Accessing the high-$\ell$ frontier under the Reduced Shear Approximation \\with $k$-cut Cosmic Shear}

\author{Anurag C. Deshpande}
\email{anurag.deshpande.18@ucl.ac.uk}
\affiliation{%
Mullard Space Science Laboratory, University College London, Holmbury St. Mary, Dorking, Surrey, RH5 6NT, UK}
\author{Peter L. Taylor}
\affiliation{%
Jet Propulsion Laboratory, California Institute of Technology, 4800 Oak Grove Drive, Pasadena, CA, 91109, USA}
\author{Thomas D. Kitching}%
\affiliation{%
Mullard Space Science Laboratory, University College London, Holmbury St. Mary, Dorking, Surrey, RH5 6NT, UK}%

\date{\today}

\begin{abstract}
The precision of Stage IV cosmic shear surveys will enable us to probe smaller physical scales than ever before, however, model uncertainties from baryonic physics and non-linear structure formation will become a significant concern. The $k$-cut method -- applying a redshift-dependent $\ell$-cut after making the Bernardeau-Nishimichi-Taruya transform -- can reduce sensitivity to baryonic physics; allowing Stage IV surveys to include information from increasingly higher $\ell$-modes. Here we address the question of whether it can also mitigate the impact of making the reduced shear approximation; which is also important in the high-$\kappa$, small-scale regime. The standard procedure for relaxing this approximation requires the repeated evaluation of the convergence bispectrum, and consequently can be prohibitively computationally expensive when included in Monte Carlo analyses. We find that the $k$-cut cosmic shear procedure suppresses the $w_0w_a$CDM cosmological parameter biases expected from the reduced shear approximation for Stage IV experiments, when $\ell$-modes up to $5000$ are probed. The maximum cut required for biases from the reduced shear approximation to be below the threshold of significance is at $k = 5.37 \, h{\rm Mpc}^{-1}$. With this cut, the predicted $1\sigma$ constraints increase, relative to the case where the correction is directly computed, by less than $10\%$ for all parameters. This represents a significant improvement in constraints compared to the more conservative case where only $\ell$-modes up to 1500 are probed \cite{ISTFpap}, and no $k$-cut is used. We also repeat this analysis for a hypothetical, comparable kinematic weak lensing survey. The key parts of code used for this analysis are made publicly available\footnote{\url{https://github.com/desh1701/k-cut_reduced_shear}}.
\end{abstract}

\maketitle


\section{\label{sec:intro}Introduction}

Cosmic shear -- the distortion of the observed ellipticities of distant galaxies resulting from weak gravitational lensing by the large-scale structure of the Universe (LSS) -- is a powerful tool to better constrain our knowledge of dark energy \cite{DETFrep}. Current weak lensing surveys \cite{DESpap, cfhtmain, kids1000} perform precision cosmology competitive with contemporary Cosmic Microwave Background surveys. Upcoming Stage IV \cite{DETFrep} cosmic shear surveys such as \emph{Euclid}\footnote{\url{https://www.euclid-ec.org/}} \cite{EuclidRB}, the \emph{Nancy Grace Roman Space Telescope}\footnote{\url{https://roman.gsfc.nasa.gov/}} \cite{WFIRSTpap}, and the Rubin Observatory\footnote{\url{https://www.lsst.org/}} \cite{LSSTpap} will offer greater than an order-of-magnitude leap in precision over the current-generation surveys \cite{SellentinStarck19}. Additionally, they will be able to probe smaller scales than previously possible (see e.g. \cite{ISTFpap}).

As a result of these improvements, we face new challenges. One such issue is the small scale sensitivity problem. This refers to the fact that the cosmic shear signal is sensitive to poorly understood physics at scales below $k = 7 \, h{\rm Mpc}^{-1}$ \cite{ssppap}. Nulling has previously been suggested as a potential solution \cite{Huterer05}. An approach that has shown promise in addressing this issue is to first apply the Bernardeau-Nishimichi-Taruya (BNT) nulling scheme \cite{bntpap}, and then take a redshift-dependent angular scale cut. This technique is known as $k$-cut cosmic shear \cite{kcutpap}.

Using $k$-cut shear to alleviate the small scale sensitivity problem, we can push our analyses to include smaller and smaller angular scales. For example, an appropriate $k$-cut would allow us to readily achieve the `optimistic' case for a \emph{Euclid}-like survey; where e.g. the inclusion of angular wave numbers of up to $\ell = 5000$ \cite{ISTFpap} would be achievable. However, at these scales, two theoretical assumptions cease to be valid; the reduced shear approximation, and the assumption that magnification bias can be neglected \cite{Deshpap}. The latter of these is a selection effect, and could potentially be addressed via a process like metacalibration \cite{metacal1, metacal2}, in particular a `selection response'. On the other hand, relaxing the reduced shear approximation requires the explicit calculation of the convergence bispectrum, which could be prohibitively computationally expensive for Stage IV experiments \cite{Deshpap} and requires a theoretical expression for the poorly understood matter bispectrum, including baryonic feedback. In this work, we demonstrate how the $k$-cut method preserves the reduced shear approximation for a Stage IV survey even at high-$\ell$. Specifically, we examine the case of a \emph{Euclid}-like experiment, as forecasting specifications for such a survey are readily available \cite{ISTFpap}. This procedure bypasses the need for the expensive computation of three-point terms, at the price of weakening cosmological parameter constraints. We also repeat this analysis for a hypothetical Tully-Fisher kinematic weak lensing survey \cite{kinematicpap, kinfirstmeasure}.

This work is structured as follows: we begin by presenting the theoretical formalism, in Section \ref{sec:theory}. We first review the standard, first-order cosmic shear power spectrum calculation; including the contribution of non-cosmological signals from the intrinsic alignments of galaxies (IA) and shot noise. Then, we discuss the formalism for relaxing the reduced shear approximation, as well as giving an overview of the BNT transform and $k$-cut cosmic shear. The Fisher matrix formalism, used to predict the cosmological parameter constraints that will be inferred from upcoming experiments, is then detailed. In Section \ref{sec:method}, we explain our modelling specifics and our choice of fiducial cosmology. Lastly, in Section \ref{sec:results}, we present our results. We compare the cosmological parameter biases resulting from making the reduced shear approximation for two different matter bispectrum models; showing that the correction calculation is robust to the choice of model. Using the most up-to-date of these models, we then demonstrate how a range of $k$-cuts affect the predicted cosmological parameter constraints and the biases from making the reduced shear approximation.

\section{\label{sec:theory}Theory}

In this section, we first review the standard cosmic shear angular power spectrum calculation. Contributions from IAs and shot noise are also described. Then, we explain how the reduced shear approximation can be relaxed. Next, we detail the BNT nulling scheme and $k$-cut cosmic shear procedure. Finally, the Fisher matrix formalism is outlined.

\subsection{\label{subsec:angpowfo}The First-order Cosmic Shear Power Spectrum}

Weak lensing distorts the observed ellipticity of distant galaxies. This change is dependent on the quantity known as reduced shear, $g$:
\begin{equation}
    \label{eq:redshear}
    g^\alpha(\boldsymbol{\theta})= \frac{\gamma^\alpha(\boldsymbol{\theta})}{1-\kappa(\boldsymbol{\theta})},
\end{equation}
where $\boldsymbol{\theta}$ is the source's position on the sky, $\gamma$ is the shear, a spin-2 quantity with index $\alpha$, and $\kappa$ is the convergence. Shear is the component of weak lensing which causes the anisotropic stretching that makes circular distributions of light elliptical, and convergence is the isotropic increase or decrease in the size of the image. In the weak lensing regime, $|\kappa|\ll 1$, so it is standard procedure to make the reduced shear approximation:
\begin{equation}
    \label{eq:RSA}
    g^\alpha(\boldsymbol{\theta}) \approx \gamma^\alpha(\boldsymbol{\theta}).
\end{equation}

The convergence in tomographic redshift bin $i$ is given by:
\begin{equation}
    \label{eq:convergence}
    \kappa_i(\boldsymbol{\theta})=\int_{0}^{\chi_{\rm lim}} {\rm d}\chi\:\delta[S_{\rm K}(\chi)\boldsymbol{\theta},\, \chi]\:W_i(\chi).
\end{equation}
It is a projection of the density contrast of the Universe, $\delta$, along the line-of-sight over comoving distance, $\chi$, to the survey's limiting comoving distance, $\chi_{\rm lim}$.
The function $S_{\rm K}(\chi)$ in equation (\ref{eq:convergence}) accounts for the curvature of the Universe, $K$, such that:
\begin{equation}
    \label{eq:SK}
    S_{\rm K}(\chi) = \begin{cases}
    |K|^{-1/2}\sin(|K|^{-1/2}\chi) & \text{\small{$K>0$ (Closed)}}\\
    \chi & \text{\small{$K=0$ (Flat)}}\\
    |K|^{-1/2}\sinh(|K|^{-1/2}\chi) & \text{\small{$K<0$ (Open)}.}
  \end{cases}
\end{equation}
$W_i$ denotes the lensing kernel for tomographic bin $i$, which is defined as follows:
\begin{align}
    \label{eq:Wi}
    W_i(\chi) &= \frac{3}{2}\Omega_{\rm m}\frac{H_0^2}{c^2}\frac{S_{\rm K}(\chi)}{a(\chi)}\int_{\chi}^{\chi_{\rm lim}}{\rm d}\chi'\:n_i(\chi')\nonumber\\
    &\times\frac{S_{\rm K}(\chi'-\chi)}{S_{\rm K}(\chi')},
\end{align}
where $\Omega_{\rm m}$ is the dimensionless present-day matter density parameter of the Universe, $H_0$ is the Hubble constant, $c$ is the speed of light in a vacuum, $a(\chi)$ is the scale factor of the Universe, and $n_i(\chi)$ is the probability distribution of galaxies within bin $i$.

Under the flat-sky approximation \cite{limitsofshear17}, the spin-2 shear is related to the convergence via:
\begin{equation}
    \label{eq:fourier}
    \widetilde{\gamma}_i^\alpha(\boldsymbol{\ell})= T^\alpha(\boldsymbol{\ell})\,\widetilde{\kappa}_i(\boldsymbol{\ell}),
\end{equation}
where $\boldsymbol{\ell}$ is the Fourier conjugate of $\boldsymbol{\theta}$, we make the `prefactor unity' approximation \cite{limitsofshear17}, and $T^\alpha(\boldsymbol{\ell})$ are trigonometric weighting functions:
\begin{align}
    \label{eq:Trigfunc1}
    T^1(\boldsymbol{\ell}) &= \cos(2\phi_\ell),\\
    \label{eq:Trigfunc2}
    T^2(\boldsymbol{\ell}) &= \sin(2\phi_\ell),
\end{align}
in which the vector $\boldsymbol{\ell}$ has angular component $\phi_\ell$, and magnitude $\ell$.

For an arbitrary shear field, two linear combinations of the shear components can be constructed: a curl-free $E$-mode, and a divergence-free $B$-mode:
\begin{align}
    \label{eq:Emode}
    \widetilde{E}_i(\boldsymbol{\ell})&=\sum_\alpha T^\alpha\:\widetilde{\gamma}_i^\alpha(\boldsymbol{\ell}),\\
    \label{eq:Bmode}
    \widetilde{B}_i(\boldsymbol{\ell})&=\sum_\alpha \sum_\beta \varepsilon^{\alpha\beta}\,T^\alpha(\boldsymbol{\ell})\:\widetilde{\gamma}_i^\beta(\boldsymbol{\ell}),
\end{align}
where $\varepsilon^{\alpha\beta}$ is the two-dimensional Levi-Civita symbol, with $\varepsilon^{12}=-\varepsilon^{21}=1$ and $\varepsilon^{11}=\varepsilon^{22}=0$. The $B$-mode of equation (\ref{eq:Bmode}) vanishes in the absence of higher-order systematic effects. This leaves the $E$-mode, for which we can define auto and cross-correlation power spectra, $C_{\ell;ij}^{\gamma\gamma}$:
\begin{equation}
    \label{eq:powerspecdef}
    \left<\widetilde{E}_i(\boldsymbol{\ell})\widetilde{E}_j(\boldsymbol{\ell'})\right> = (2\pi)^2\,\delta_{\rm D}^2(\boldsymbol{\ell}+\boldsymbol{\ell'})\,C_{\ell;ij}^{\gamma\gamma},
\end{equation}
with $\delta_{\rm D}^2$ being the two-dimensional Dirac delta. Under the assumption of the Limber approximation, where only $\ell$-modes in the plane of the sky are taken to contribute to the lensing signal, the power spectra themselves are:
\begin{equation}
    \label{eq:Cl}
    C_{\ell;ij}^{\gamma\gamma} = \int_0^{\chi_{\rm lim}}{\rm d}\chi\frac{W_i(\chi)W_j(\chi)}{S^{\,2}_{\rm K}(\chi)}P_{\delta\delta}(k, \chi),
\end{equation}
where $P_{\delta\delta}(k, \chi)$ is the matter power spectrum. Comprehensive reviews of this standard calculation can be found in \cite{Kilbinger15, Munshirev}.

\subsection{\label{subsec:IAs}Intrinsic Alignments and Shot Noise}

In reality, the angular power spectrum measured from galaxy surveys contains non-cosmological signals, in addition to the cosmic shear contribution from equation (\ref{eq:Cl}). One such component is the result of the IA of galaxies \cite{JoachimiIAs}. Galaxies that form in similar tidal environments have preferred, intrinsically correlated, alignments. The observed ellipticity of a galaxy, $\epsilon$ can be described to first-order as:
\begin{equation}
    \label{eq:galelip}
    \epsilon = \gamma + \gamma^{\rm I} + \epsilon^s,
\end{equation}
where $\gamma$ is from cosmic shear, $\gamma^{\rm I}$ is the IA contribution, and $\epsilon^s$ is the galaxy's source ellipticity in the absence of any IA. A theoretical two-point statistic (e.g. the angular power spectrum) calculated from equation (\ref{eq:galelip}) would then consist of four kinds of terms: $\langle\gamma\gamma\rangle,\langle\gamma^{\rm I}\gamma\rangle$, $\langle\gamma^{\rm I}\gamma^{\rm I}\rangle$, and a shot noise term from the uncorrelated part of the unlensed source ellipticities, $\epsilon^s$.

Accordingly, the observed angular power spectra, $C_{\ell;ij}^{\epsilon\epsilon}$, contain contributions from all these terms:
\begin{equation}
    \label{eq:ObsCl}
    C_{\ell;ij}^{\epsilon\epsilon} = C_{\ell;ij}^{\gamma\gamma} + C_{\ell;ij}^{{\rm I}\gamma} + C_{\ell;ij}^{\gamma{\rm I}} + C_{\ell;ij}^{\rm II} + N_{\ell;ij}^\epsilon,
\end{equation}
where $C_{\ell;ij}^{\gamma\gamma}$ are the cosmic shear spectra of equation (\ref{eq:Cl}), $C_{\ell;ij}^{{\rm I}\gamma}$ represent the correlation between the background shear and the foreground IA, $C_{\ell;ij}^{\gamma{\rm I}}$ are the correlation of the foreground shear with background IA, $C_{\ell;ij}^{\rm II}$ are the auto-correlation spectra of the IAs, and $N_{\ell;ij}^\epsilon$ is the shot noise. The $C_{\ell;ij}^{\gamma{\rm I}}$ spectra are zero except in the case of when photometric redshifts cause scattering of observed redshifts between bins.

The additional non-zero IA spectra can be described in an analogous manner to the shear power spectra, through the use of the non-linear alignment (NLA) model \citep{NLAmodel}:
\begin{align}
    \label{eq:cllig}
    C_{\ell;ij}^{{\rm I}\gamma} &= \int_0^{\chi_{\rm lim}}\frac{{\rm d}\chi}{S^{\,2}_{\rm K}(\chi)}[W_i(\chi)n_j(\chi)+n_i(\chi)W_j(\chi)]\nonumber\\
    &\times P_{\delta {\rm I}}(k, \chi),\\
    \label{eq:clli}
    C_{\ell;ij}^{\rm II} &= \int_0^{\chi_{\rm lim}}\frac{{\rm d}\chi}{S^{\,2}_{\rm K}(\chi)}n_i(\chi)n_j(\chi)\,P_{\rm II}(k, \chi),
\end{align}
where $P_{\delta {\rm I}}(k, \chi)$ and $P_{\rm II}(k, \chi)$ are the IA power spectra, and are defined as functions of the matter power spectra:
\begin{align}
    \label{eq:pdi}
    P_{\delta {\rm I}}(k, \chi) &= \bigg(-\frac{\mathcal{A}_{\rm IA}\mathcal{C}_{\rm IA}\Omega_{\rm m}}{D(\chi)}\bigg)\:\:P_{\delta\delta}(k,\chi),\\
    \label{eq:pii}
    P_{\rm II}(k, \chi) &= \bigg(-\frac{\mathcal{A}_{\rm IA}\mathcal{C}_{\rm IA}\Omega_{\rm m}}{D(\chi)}\bigg)^2P_{\delta\delta}(k,\chi).
\end{align}
Within these equations, $\mathcal{A}_{\rm IA}$ and $\mathcal{C}_{\rm IA}$ are free model parameters to be determined by fitting to data or simulations, and $D(\chi)$ is the growth factor of density perturbations in the Universe, as a function of comoving distance.

The shot noise, which is the last of the terms in equation (\ref{eq:ObsCl}), is written as:
\begin{align}
    \label{eq:shotnoise}
    N_{\ell;ij}^\epsilon = \frac{\sigma_\epsilon^2}{\bar{n}_{\rm g}/N_{\rm bin}}\delta_{ij}^{\rm K},
\end{align}
where $\sigma_\epsilon^2$ is the variance of the observed ellipticities in the galaxy sample, $\bar{n}_{\rm g}$ is the galaxy surface density of the survey, $N_{\rm bin}$ is the number of tomographic bins used, and $\delta_{ij}^{\rm K}$ is the Kronecker delta. The shot noise term is zero for cross-correlation spectra because the ellipticities of galaxies at different comoving distances should be uncorrelated. Equation (\ref{eq:shotnoise}) assumes that the tomographic bins used in the survey are equi-populated.

\subsection{\label{subsec:RScor}Relaxing the Reduced Shear Approximation}

The procedure for relaxing the reduced shear approximation involves Taylor expanding equation (\ref{eq:redshear}) around $\kappa=0$, and retaining terms up to and including second-order: \cite{Deshpap, KrauseHirata, Shapiro09}:
\begin{equation}
    \label{eq:gexpan}
    g^\alpha(\boldsymbol{\theta})=\gamma^\alpha(\boldsymbol{\theta})+(\gamma^\alpha\kappa)(\boldsymbol{\theta})+\mathcal{O}(\kappa^3).
\end{equation}
This expression for $g^\alpha$ is then substituted for $\gamma^\alpha$ in equation (\ref{eq:Emode}). Recomputing the power spectrum, we recover equation (\ref{eq:Cl}) plus a second-order correction term:
\begin{align}
    \label{eq:dCl}
    \delta C^{\rm RS}_{\ell;ij} &= \int_0^\infty\frac{{\rm d}^2\boldsymbol{\ell'}}{(2\pi)^2}\cos(2\phi_{\ell'}-2\phi_\ell)\nonumber\\
    &\times B_{ij}^{\kappa\kappa\kappa}(\boldsymbol{\ell}, \boldsymbol{\ell'}, -\boldsymbol{\ell}-\boldsymbol{\ell'}),
\end{align}
in which $B_{ij}^{\kappa\kappa\kappa}$ are the two-redshift convergence bispectra. Under the assumption of an isotropic universe, we are always free to set $\phi_{\ell}=0$.

The convergence bispectra can then be safely expressed subject to the Limber approximation \cite{postlimbrs} as projections of the matter bispectra, $B_{\delta\delta\delta}$:
\begin{align}
    \label{eq:bispecK}
    B_{ij}^{\kappa\kappa\kappa}(\boldsymbol{\ell_1}, \boldsymbol{\ell_2}, \boldsymbol{\ell_3}) &= B_{iij}^{\kappa\kappa\kappa}(\boldsymbol{\ell_1}, \boldsymbol{\ell_2}, \boldsymbol{\ell_3}) + B_{ijj}^{\kappa\kappa\kappa}(\boldsymbol{\ell_1}, \boldsymbol{\ell_2}, \boldsymbol{\ell_3})\nonumber\\ &=\int_0^{\chi_{\rm lim}}\frac{{\rm d}\chi}{S^{\,4}_{\rm K}(\chi)}W_i(\chi)W_j(\chi)\nonumber\\
    &\times [W_i(\chi)+W_j(\chi)] \nonumber\\
    &\times B_{\delta\delta\delta}(\boldsymbol{k_1},\boldsymbol{k_2},\boldsymbol{k_3},\chi).
\end{align}
The analytic form of the matter bispectrum is not fully known. Instead, expressions are typically obtained by fitting to N-body simulations \cite{ScocCouch, Gilmarin, bihalofit}. In this work, we examine two such approaches. 

The first approach starts from second-order perturbation theory \cite{Frypap}, and then fits the resulting expression to simulations. We denote this approach by SC, after the first work to propose this methodology \cite{ScocCouch}. Here, the matter bispectrum can be written as:
\begin{align}
    \label{eq:bispecSC}
    B_{\delta\delta\delta}(\boldsymbol{k_1},\boldsymbol{k_2},\boldsymbol{k_3},\chi) &= 2F_2^{\rm eff}\,(\boldsymbol{k_1},\boldsymbol{k_2})\,P_{\delta\delta}(k_1, \chi)P_{\delta\delta}(k_2, \chi) \nonumber\\
    &+ \text{cyc. perms.} ,
\end{align}
with:
\begin{align}
\label{eq:Feff}
        F_2^{\rm eff}(\boldsymbol{k_1},\boldsymbol{k_2}) &= \frac{5}{7}\,a(n_{\rm s},k_1)\,a(n_{\rm s},k_2) \nonumber\\
        &+ \frac{1}{2}\frac{\boldsymbol{k_1}\cdot\boldsymbol{k_2}}{k_1k_2}\,\bigg(\frac{k_1}{k_2}+\frac{k_2}{k_1}\bigg)\,b(n_{\rm s},k_1)\,b(n_{\rm s},k_2) \nonumber\\
        &+\frac{2}{7}\,\bigg(\frac{\boldsymbol{k_1}\cdot\boldsymbol{k_2}}{k_1k_2}\bigg)^2c(n_{\rm s},k_1)\,c(n_{\rm s},k_2),
\end{align}
where $a, b$, and $c$ are fitting functions given in \cite{ScocCouch}.

A more contemporary approach adopts the \texttt{Halofit} formalism \cite{Takahashi12} for the matter power spectrum, to also describe the matter bispectrum \cite{bihalofit}. We denote this approach by BH, as this technique is known as \texttt{BiHalofit}. In this paradigm, the matter bispectrum constitutes one-halo (1h), and three-halo (3h) terms:
\begin{align}
    \label{eq:bispecBF}
    B_{\delta\delta\delta}(\boldsymbol{k_1},\boldsymbol{k_2},\boldsymbol{k_3},\chi) &= B_{1h}(\boldsymbol{k_1},\boldsymbol{k_2},\boldsymbol{k_3},\chi)\nonumber \\
    &+ B_{3h}(\boldsymbol{k_1},\boldsymbol{k_2},\boldsymbol{k_3},\chi).
\end{align}
These terms are then determined by fitting to N-body simulations. A full description of these can be found in Appendix B of \cite{bihalofit}. 

\subsection{\label{subsec: kcut}$k$-cut Cosmic Shear}

Given that the shear angular power spectrum is a projection of the matter power spectrum, to remove sensitivity to physical scales below a certain k-mode we must remove angular scales above the corresponding $\ell$-mode. One may imagine that, in the regime of the Limber approximation, this could simply involve removing information where $\ell > k\chi$. However, in reality lensing kernels are broad; meaning that lenses across a range of distances and scales contribute power to the same $\ell$-mode. Consequently, this simple method of removing scales is not effective on its own \cite{ssppap}.

A solution comes in the form of the BNT nulling scheme \cite{bntpap}. In this formalism, the observed tomographic angular power spectrum can be re-weighted in such a way that each redshift bin retains only the information about lenses within a small redshift range. This procedure can be illustrated by first considering three discrete source planes. Then, the BNT weighted convergence, assuming flatness, can be written as:
\begin{align}
    \label{eq:bntconv}
    \kappa^{\rm BNT}=\frac{3 \Omega_{m} H_{0}^{2}}{2 c^{2}} \int_{0}^{\chi_{\beta}} \mathrm{d} r \frac{\delta(\chi)}{a(\chi)} w(\chi),
\end{align}
where $\chi_\beta$ is the comoving distance to source plane $i$, and:
\begin{align}
    \label{eq:smallw}
    w(\chi)=\sum_{\beta, \chi_{\beta}>\chi} p_{\beta} \frac{\chi_{\beta}-\chi}{\chi_{\beta}},
\end{align}
where $p_\beta$ are the weights for planes $\beta = \{1,2,3\}$ with $\chi_1 < \chi_2 < \chi_3$, for the three bin case. In the BNT scheme, weights are then chosen such that $w(\chi < \chi_1) = 0$. This coupled with the fact that lenses with $\chi > \chi_3$ will not contribute to the re-weighted convergence, means that $\kappa^{\rm BNT}$ will only be sensitive to lenses with comoving distances in the range $\chi_1 \leq \chi < \chi_3$. This argument can be generalized \cite{xcutpaper} for an arbitrary number of continuous source bins; leading to the construction of a weighting matrix, $\boldsymbol{M}$, that can be applied to the observed tomographic angular power spectra: 
\begin{align}
    \label{eq:bntcl}
   \boldsymbol{C}_{\ell}^{\rm BNT} = \boldsymbol{M} \boldsymbol{C}_{\ell} \boldsymbol{M}^T,
\end{align}
where $\boldsymbol{C}_{\ell}$ is a matrix of the $C_{\ell; ij}$ for all tomographic bin combinations, at the given $\ell$-mode, and $\boldsymbol{C}^{\rm BNT}_{\ell}$ is its BNT-nulled counterpart.

For a given $k$-cut, we remove information where $\ell>k_{\rm cut} \chi_{i}^{\rm mean}$ from the tomographic BNT-nulled angular power spectrum of bin $i$. Here, we use the mean comoving distance of the redshift bin rather than the minimum distance to the bin in order to avoid removing the first bin entirely. This has negligible impact on reduction in sensitivity to small scales \cite{kcutpap}. 

\subsection{\label{subsec:fisher} Fisher Matrices and Bias Formalism}

The cosmological parameter constraints for a given survey can be predicted by using the Fisher matrix formalism \citep{Tegmark97}. The Fisher matrix is given by the expectation of the Hessian of the likelihood.
By safely assuming a Gaussian likelihood \cite{LSSTnongauss, gausspeter}, we can rewrite the Fisher matrix in terms of only the mean of the data vector, and the covariance of the data. For the cosmic shear case, we note that the mean of the shear field is zero. Under the Gaussian covariance assumption, the specific Fisher matrix for a cosmic shear survey is then (see e.g. \cite{ISTFpap} for a detailed derivation):
\begin{align}
    \label{eq:fishshear}
    F_{\tau \zeta}&=f_{\mathrm{sky}} \sum_{\ell=\ell_{\min }}^{\ell_{\max }} \Delta \ell\left(\ell+\frac{1}{2}\right) \nonumber\\
    &\times\operatorname{tr}\left[\frac{\partial \boldsymbol{C}_{\ell}}{\partial \theta_{\tau}} {\boldsymbol{C}_{\ell}}^{-1} \frac{\partial \boldsymbol{C}_{\ell}}{\partial \theta_{\zeta}} {\boldsymbol{C}_{\ell}}^{-1}\right],
\end{align}
where $f_{\rm sky}$ is the fraction of sky included in the survey, $\Delta \ell$ is the bandwidth of $\ell$-modes sampled, the sum is over these blocks in $\ell$, and $\tau$ and $\zeta$ refer to parameters of interest, $\theta_\tau$ and $\theta_\zeta$. The predicted uncertainty for a parameter, $\tau$, is then calculated with:
\begin{align}
    \label{eq:sigfish}
    \sigma_\tau = \sqrt{{F_{\tau\tau}}^{-1}}.
\end{align}

This formalism can be adapted to show how biased the predicted cosmological parameter values will be when a systematic effect in the data is neglected \cite{Fishbias}:
\begin{align}
    \label{eq:bias}
    b\left(\theta_{\tau}\right) &=\sum_{\zeta}F^{-1}_{\tau\zeta} f_{\mathrm{sky}} \sum_{\ell} \Delta \ell\left(\ell+\frac{1}{2}\right) \nonumber\\
    &\times  \operatorname{tr}\left[\delta \boldsymbol{C}_{\ell} \, {\boldsymbol{C}_{\ell}}^{-1} \frac{\partial \boldsymbol{C}_{\ell}}{\partial \theta_{\zeta}} {\boldsymbol{C}_{\ell}}^{-1}\right],
\end{align}
where $\delta \boldsymbol{C}_\ell$ is a matrix with every tomographic bin combination of the systematic effect, $\delta C_{\ell; ij}$, evaluated at mode $\ell$. In this work, these systematic effect terms are given by the reduced shear correction of equation (\ref{eq:dCl}).

\section{\label{sec:method}Methodology}

In order to examine whether $k$-cut cosmic shear can be used to minimise the impact of the reduced shear approximation on Stage IV surveys, we adopt forecasting specifications for a \emph{Euclid}-like survey \cite{ISTFpap}. The $k$-cut technique enables the inclusion of information from smaller angular scales, making the `optimistic' scenario for such a survey, where $\ell$-modes up to 5000 are studied, more achievable. Accordingly, we compute the reduced shear correction, and carry out the corresponding $k$-cut analysis, up to this maximum $\ell$. This is compared to the `pessimistic' case for such an experiment where only $\ell$-modes up to 1500 are included, and no $k$-cut is necessary \cite{ISTFpap}.

The fraction of sky that will be covered by a \emph{Euclid}-like survey is $f_{\rm sky} = 0.36$. The intrinsic variance of unlensed galaxy ellipticities is modelled with two components, each of value 0.21, so that the root-mean-square intrinsic ellipticity is $\sigma_\epsilon = \sqrt{2}\times0.21 \approx 0.3$. The surface density of galaxies will be $\bar{n}_{\rm g}=30$ arcmin$^{-2}$. We examine the case where the data consists of ten equi-populated redshift bins with limits: \{0.001, 0.418, 0.560, 0.678, 0.789, 0.900, 1.019, 1.155, 1.324, 1.576, 2.50\}.

The galaxy distributions within these tomographic bins, assuming they are determined with photometric redshift estimates, are modelled according to:
\begin{equation}
    \label{eq:ncfht}
    {\mathcal N}_i(z) = \frac{\int_{z_i^-}^{z_i^+}{\rm d}z_{\rm p}\,\mathfrak{n}(z)p_{\rm ph}(z_{\rm p}|z)}{\int_{z_{\rm min}}^{z_{\rm max}}{\rm d}z\int_{z_i^-}^{z_i^+}{\rm d}z_{\rm p}\,\mathfrak{n}(z)p_{\rm ph}(z_{\rm p}|z)},
\end{equation}
where $z_{\rm p}$ is measured photometric redshift, $z_i^-$ and $z_i^+$ are edges of the $i$-th redshift bin, $z_{\rm min}$ and $z_{\rm max}$ define the range of redshifts covered by the survey, and $\mathfrak{n}(z)$ is the true distribution of galaxies with redshift, $z$, which is defined using the expression \cite{EuclidRB}:
\begin{equation}
    \label{eq:ntrue}
    \mathfrak{n}(z) \propto \bigg(\frac{z}{z_0}\bigg)^2\,{\rm exp}\bigg[-\bigg(\frac{z}{z_0}\bigg)^{3/2}\bigg],
\end{equation}
wherein $z_0=z_{\rm m}/\sqrt{2}$, with $z_{\rm m}=0.9$ being the median redshift of the survey. The function $p_{\rm ph}(z_{\rm p}|z)$ exists to account for the probability that a galaxy at redshift $z$ is measured to have a redshift $z_{\rm p}$, and is given by:
\begin{align}
\label{eq:pphot}
        p_{\rm ph}(z_{\rm p}|z) &= \frac{1-f_{\rm out}}{\sqrt{2\pi}\sigma_{\rm b}(1+z)}\,{\rm exp}\Bigg\{-\frac{1}{2}\bigg[\frac{z-c_{\rm b}z_{\rm p}-z_{\rm b}}{\sigma_{\rm b}(1+z)}\bigg]^2\Bigg\} \nonumber\\
        &+ \frac{f_{\rm out}}{\sqrt{2\pi}\sigma_{\rm o}(1+z)}\nonumber\\
        &\times{\rm exp}\Bigg\{-\frac{1}{2}\bigg[\frac{z-c_{\rm o}z_{\rm p}-z_{\rm o}}{\sigma_{\rm o}(1+z)}\bigg]^2\Bigg\}.
\end{align}
In this equation, the first term on the right-hand side describes the multiplicative and additive bias in redshift determination for the fraction of sources with a well measured redshift, while the second term accounts for the effect of a fraction of catastrophic outliers, $f_{\rm out}$. The values assigned to the parameters in this equation are stated in Table \ref{tab:phphotparams}. Then, the galaxy distribution as a function of comoving distance is $n_i(\chi) = {\mathcal N}_i(z){\rm d}z/{\rm d}\chi$.
\begin{table}[t]
    \centering
    \caption{Parameter values in this investigation in order to describe the probability distribution function of the photometric redshift distribution of sources, in equation (\ref{eq:pphot}).}
    \begin{tabular}{c c}
    \hline\hline
    Parameter & Value\\
    \hline
    $c_{\rm b}$ & 1.0\\
    $z_{\rm b}$ & 0.0\\
    $\sigma_{\rm b}$ & 0.05\\
    $c_{\rm o}$ & 1.0\\
    $z_{\rm o}$ & 0.1\\
    $\sigma_{\rm o}$ & 0.05\\
    $f_{\rm out}$ & 0.1\\
    \hline\hline
    \end{tabular}
    \label{tab:phphotparams}
\end{table}

Kinematic lensing has been proposed as a method to reduce shape noise in weak lensing by an order of magnitude \cite{kinematicpap}. It is predicated on spectroscopic measurements of disk galaxy rotation and use of the Tully-Fisher relation in order to control for the intrinsic orientations of galaxy disks. Here, we study the effect of $k$-cut cosmic shear on the hypothetical TF-Stage III survey described in \cite{kinematicpap}. This survey includes $\ell$-modes up to 5000, has $f_{\rm sky}=0.12$, with an intrinsic ellipticity of $\sigma_\epsilon = 0.021$, and a surface density of galaxies of $\bar{n}_{\rm g}=1.1$ arcmin$^{-2}$. We consider the survey to have ten equi-populated redshift bins with limits: $\{$0.001, 0.568, 0.654, 0.723, 0.788, 0.851, 0.921, 0.999, 1.097, 1.243, 1.68$\}$. A kinematic survey will not have IA contributions. The galaxy distribution for such a survey is modelled by:
\begin{align}
    \label{eq:kindist}
    \mathcal{N}_i(z) \propto z^{\alpha} e^{-\left(\frac{z}{z_{0}}\right)^{\beta}},
\end{align}
with $\alpha=29.98$, $z_0=1.10\times10^{-6}$, and $\beta=0.33$.

This work assumes a flat $w_0w_a$CDM fiducial cosmology. Allowing for a time-varying dark energy equation-of-state, the model consists of the following parameters: the present-day total matter density parameter $\Omega_{\rm m}$, the present-day baryonic matter density parameter $\Omega_{\rm b}$, the Hubble parameter $h=H_0/100$km\:s$^{-1}$Mpc$^{-1}$, the spectral index $n_{\rm s}$, the RMS value of density fluctuations on 8 $h^{-1}$Mpc scales $\sigma_8$, the present-day value of the dark energy equation of state $w_0$, the high-redshift value of the dark energy equation of state $w_a$, and massive neutrinos with a sum of masses $\sum m_\nu\ne 0$. We choose the same fiducial parameter values as presented in \cite{ISTFpap}. We explicitly state these in Table \ref{tab:cosmology}. The BNT matrices are calculated using the code\footnote{\url{https://github.com/pltaylor16/x-cut}} of \cite{xcutpaper}. Additionally, to calculate the matter power spectrum, we use the publicly available \texttt{CAMB}\footnote{\url{https://camb.info/}} cosmology package \cite{cambpap}, with \texttt{Halofit} \cite{Takahashi12} and corrections from \cite{MeadHF} used to compute the non-linear contributions. Comoving distances are computed with \texttt{Astropy}\footnote{\url{http://www.astropy.org}} \cite{astropy1, astropy2}. To obtain the matter bispectrum of the BH approach, we employ the publicly available \texttt{C} code\footnote{\url{http://cosmo.phys.hirosaki-u.ac.jp/takahasi/codes_e.htm}} supplied with \cite{bihalofit}. The IA power spectra are modelled with the parameter values: $\mathcal{A}_{\rm IA}=1.72$ and $\mathcal{C}_{\rm IA}=0.0134$ \cite{ISTFpap}. Our Fisher matrix contains the following parameters: $\Omega_{\rm m}, \Omega_{\rm b}, h, n_{\rm s}, \sigma_8, w_0, w_a,$ and $\mathcal{A}_{\rm IA}$, for consistency with \cite{ISTFpap}.
\begin{table}[t]
\centering
\caption{Fiducial values of $w_0w_a$CDM cosmological parameters adopted in this work. Values were selected to match \citep{ISTFpap}.}
\label{tab:cosmology}
\begin{tabular}{c c}
\hline\hline
Cosmological Parameter & Fiducial Value\\
\hline
$\Omega_{\rm m}$ & 0.32 \\
$\Omega_{\rm b}$ & 0.05 \\
$h$ & 0.67 \\
$n_{\rm s}$ & 0.96 \\
$\sigma_8$ & 0.816 \\
$\sum m_\nu$ (eV) & 0.06 \\
$w_0$ & $-1$ \\
$w_a$ & 0  \\
\hline\hline
\end{tabular}
\end{table}

\section{\label{sec:results}Results and Discussion}

In this section, we demonstrate the effect $k$-cut cosmic shear has on addressing the biases resulting from the reduced shear approximation, for a \emph{Euclid}-like experiment and a hypothetical kinematic survey. We begin by comparing the cosmological parameter biases, for the standard calculation with no $k$-cut, found when the reduced shear approximation is relaxed with either the SC or BH bispectrum models. Next, the change in parameter constraints and biases for the BNT transformed power spectra with a range of $k$-cuts are shown; first for a \emph{Euclid}-like survey, and then a kinematic lensing survey.

\subsection{\label{subsec:bimodels}Comparing Matter Bispectrum Models}

\begin{figure}[t]
\centering
\includegraphics[width=1.0\linewidth]{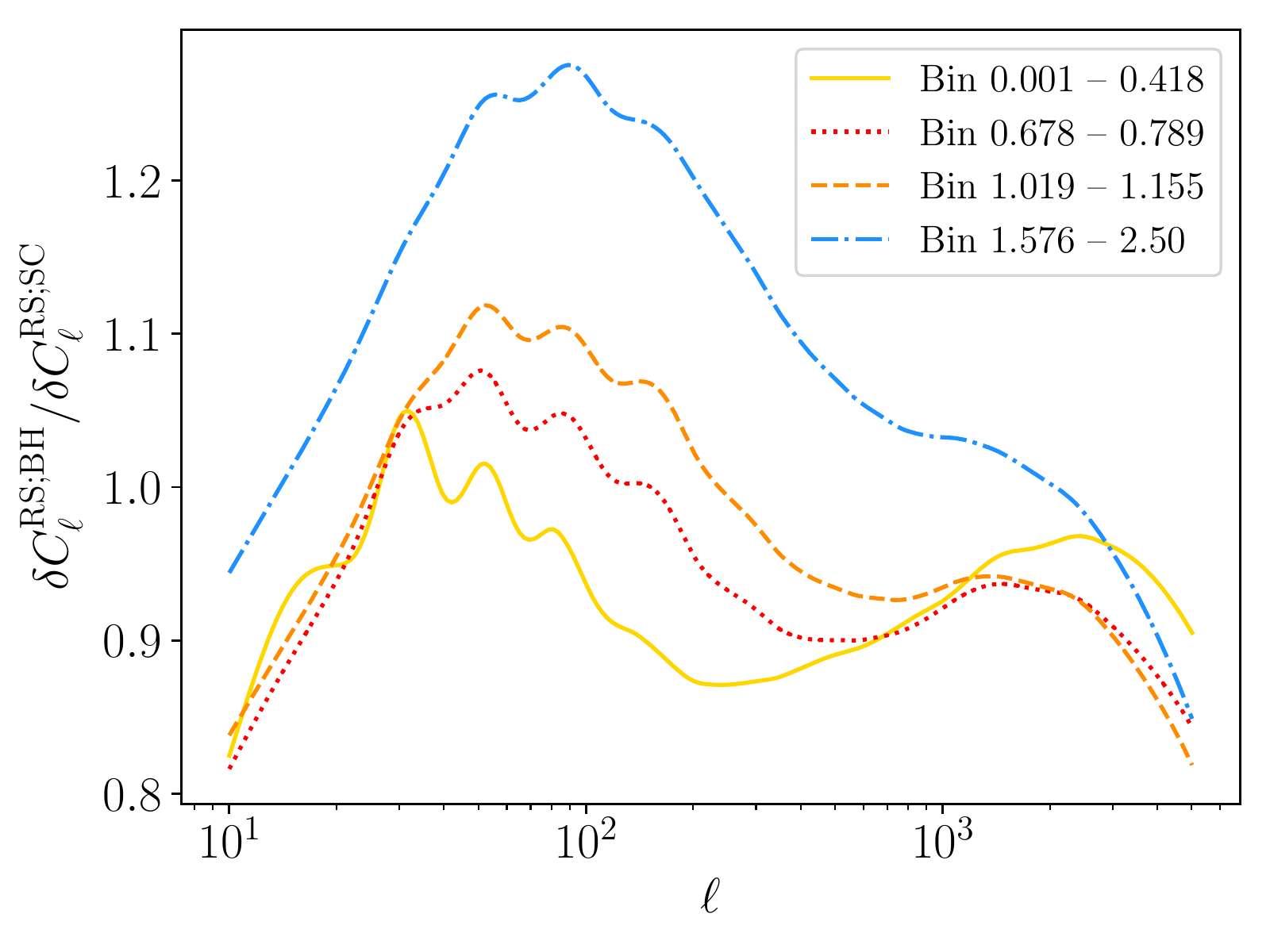}
\caption{Ratio of reduced shear corrections calculated with two different matter bispectrum models. The first of these uses the approach of \cite{ScocCouch} and is labelled by SC, whereas the second is the \texttt{BiHalofit} model \cite{bihalofit} and is denoted by BH. The correction terms for four different auto-correlation spectra across the survey's anticipated redshift range are presented, and are representative of all the spectra. The most extreme disagreement between the models occurs at $\ell = 89$, where they disagree by 27$\%$. We note that the reduced shear correction is negligible at these scales, and only becomes significant at scales above $\ell \sim 1000$ \cite{Deshpap}, at which point the two models are in closer agreement.}
\label{fig:mod_comp}
\end{figure}

The ratio of the reduced shear correction of equation (\ref{eq:dCl}) calculated using the BH bispectrum, relative to the correction calculated using the SC bispectrum is shown in Figure \ref{fig:mod_comp}. Here, the correction terms for the auto-correlation of four bins, with redshift-limits: 0.001 -- 0.418, 0.678 -- 0.789, 1.019 -- 1.155, and 1.576 -- 2.50, are shown in order to illustrate the difference between the two models. The consequent difference in the predicted parameter biases from using the two models is stated in Table \ref{tab:bi_mod_comp}.

\begin{table}[b]
\centering
\caption{Cosmological parameter biases predicted if the reduced shear correction is neglected for two different matter bispectrum models. The SC model uses the fitting formulae of \cite{ScocCouch}, while BH is the \texttt{Bihalofit} model \cite{bihalofit}. The difference between the two approaches is also stated, and is not significant. Here $\sigma$ denotes the 1$\sigma$ uncertainty.}
\label{tab:bi_mod_comp}
\begin{tabular}{c c c c}
\hline\hline
Cosmological & SC Model & BH Model & Absolute Difference in\\
Parameter & Bias/$\sigma$ & Bias/$\sigma$ & Biases/$\sigma$ \\
\hline
$\Omega_{\rm m}$ & -0.32 & -0.28 & 0.04\\
$\Omega_{\rm b}$ & -0.011 & -0.0056 & 0.0044\\
$h$ & 0.025 & 0.027 & 0.002\\
$n_{\rm s}$ & 0.14 & 0.11 & 0.03\\
$\sigma_8$ & 0.27 & 0.24 & 0.03\\
$w_0$ & -0.40 & -0.33 & 0.07\\
$w_a$ & 0.28  & 0.23 & 0.05\\
\hline\hline
\end{tabular}
\end{table}

From Figure \ref{fig:mod_comp}, we see that the two approaches produce correction terms that differ at most by 27$\%$. At low-$\ell$ and at all but the highest redshifts, the BH model produces \clearpage\newpage
\onecolumngrid

\begin{figure}[h]
\centering
\includegraphics[width=1.0\linewidth]{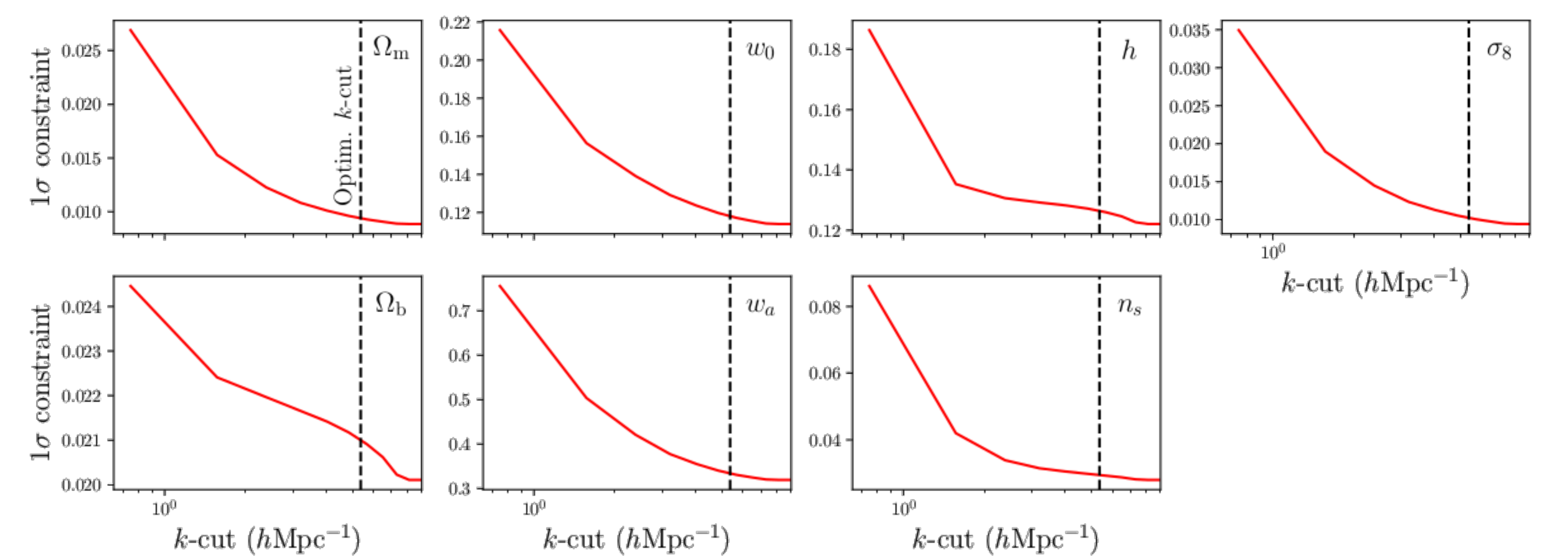}
\caption{Change in the 1$\sigma$ cosmological parameter constraints predicted for a \emph{Euclid}-like survey, when a range of $k$-cuts are applied. These results are for the `optimistic' case for such a survey, where $\ell$-modes up to 5000 are included. Unsurprisingly, the constraints weaken as lower $k$-cuts are taken; corresponding to more information being removed. The black dashed line at $k = 5.37 \, h$Mpc$^{-1}$ marks the maximum $k$-cut required for biases from the reduced shear correction to not be significant.}
\label{fig:sig_kcut}
\end{figure}
\begin{figure}[h]
\centering
\includegraphics[width=1.0\linewidth]{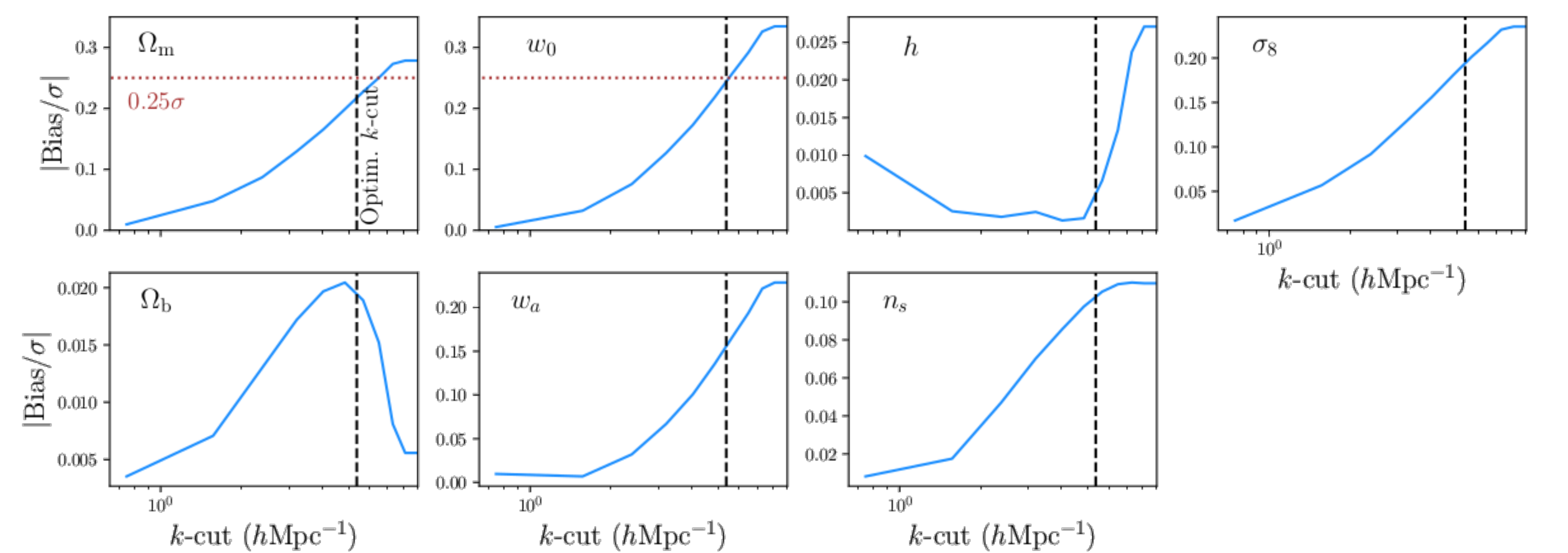}
\caption{Change in cosmological parameter biases with changing $k$-cuts, when the reduced shear correction is neglected, for a \emph{Euclid}-like survey. The values are reported as a fraction of the 1$\sigma$ uncertainty of the respective parameter. A parameter is considered to be significantly biased if the bias is greater than 0.25$\sigma$. Beyond this point, the biased and unbiased confidence regions overlap less than 90$\%$. These results are for the `optimistic' case for a \emph{Euclid}-like survey, where $\ell$-modes up to 5000 are included. The black dashed line at $k = 5.37 \, h$Mpc$^{-1}$ marks the maximum $k$-cut required for biases from the reduced shear correction to not be significant. The brown dotted line denotes the threshold for a bias to be significant. Generally, a lower $k$-cut corresponds to smaller biases, as sensitivity is reduced to regions where the reduced shear correction is largest.}
\label{fig:bias_kcut}
\end{figure}
\twocolumngrid

\noindent a correction smaller than the SC one. The BH correction then increases until the two models produce comparable results at $\ell \sim 100$. Beyond this $\ell$-mode, the BH model once again produces a smaller correction value than the SC approach. For the highest redshift bins, the same trend persists. However, in this case the corrections start off being comparable, before the BH term becomes greater than the SC correction. After peaking at scales of $\ell \sim 100$, the BH correction reduces again. The greatest differences between the two models occur at lower $\ell$-modes, where the reduced shear correction is typically negligible \cite{Deshpap}. Additionally, these differences are likely to be dwarfed by baryonic model uncertainties.

Despite these differences, Table \ref{tab:bi_mod_comp} shows that the resulting cosmological parameter biases from the two models are not significantly different. Accordingly, although the BH and SC models can differ significantly at calculating the matter bispectrum for certain scales and configurations \cite{bihalofit}, the reduced shear correction calculation can be considered robust to the choice of matter bispectrum model. For all results that follow, we use the BH matter bispectrum model. \clearpage \newpage

\onecolumngrid

\begin{table}[h]
\centering
\caption{Predicted parameter uncertainties, and biases from neglecting the reduced shear approximation, for a \emph{Euclid}-like survey under three different scenarios. The `optimistic' scenario is when $\ell$-modes up to 5000 are included, and no $k$-cut is made, while the `maximum $k$-cut' columns denote the situation where $\ell$-modes up to 5000 are included, but a $k$-cut is taken at $k=5.37$ $h$Mpc$^{-1}$, as this is the maximum $k$-cut to achieve non-significant biases. Finally, the `pessimistic' case is when only $\ell$-modes up to 1500 are included, and no $k$-cut is taken. The `maximum $k$-cut' option is able to suppress the biases to the point of not being significant, while still achieving more precise constraints than the `pessimistic' option. Here $\sigma$ denotes the 1$\sigma$ uncertainty.}
\label{tab:unctab}
\begin{tabular}{c c c c | c c c}
\hline\hline
Cosmological & Optimistic ($\ell_{\rm max}=5000$) & Maximum $k$-cut & Pessimistic ($\ell_{\rm max}=1500$) & Optimistic & Maximum $k$-cut & Pessimistic \\
Parameter & Uncertainty (1$\sigma$) & Uncertainty (1$\sigma$) & Uncertainty (1$\sigma$) & Bias/$\sigma$ & Bias/$\sigma$ & Bias/$\sigma$\\
\hline
$\Omega_{\rm m}$& 0.0089 & 0.0094 & 0.013 & -0.28 & -0.22 & -0.076\\
$\Omega_{\rm b}$& 0.020 & 0.021 & 0.022 & -0.0056 & -0.020 & -0.012\\
$h$& 0.12 & 0.13 & 0.13 & 0.027 & 0.0043 & -0.001\\
$n_{\rm s}$& 0.028 & 0.029 & 0.035 & 0.11 & 0.10 & 0.040\\
$\sigma_8$& 0.0094 & 0.010 & 0.015 & 0.24 & 0.19 & 0.083\\
$w_0$& 0.11 & 0.12 & 0.14 & -0.33 & -0.24 & -0.064\\
$w_a$& 0.32 & 0.33 & 0.44 & 0.23 & 0.15 & 0.024\\
\hline\hline
\end{tabular}
\end{table}
\begin{table}[h]
\centering
\caption{Predicted cosmological parameter constraints, and biases from neglecting the reduced shear approximation, for a TF-Stage III \cite{kinematicpap} kinematic lensing survey. Three different scenarios are presented here. The `optimistic' scenario is when $\ell$-modes up to 5000 are included, and no $k$-cut is made, while the `maximum $k$-cut' columns denote the situation where $\ell$-modes up to 5000 are included, but a $k$-cut is taken at $k=5.82$ $h$Mpc$^{-1}$, as this is the maximum $k$-cut to achieve non-significant biases. Finally, the `pessimistic' case is when only $\ell$-modes up to 1500 are included, and no $k$-cut is taken. The `maximum $k$-cut' option is able to suppress the biases to the point of not being significant, while still achieving more precise constraints than the `pessimistic' option. Here $\sigma$ denotes the 1$\sigma$ uncertainty.}
\label{tab:kintabunc}
\begin{tabular}{c c c c | c c c}
\hline\hline
Cosmological & Optimistic ($\ell_{\rm max}=5000$) & Maximum $k$-cut & Pessimistic ($\ell_{\rm max}=1500$) & Optimistic & Maximum $k$-cut & Pessimistic \\
Parameter & Uncertainty (1$\sigma$) & Uncertainty (1$\sigma$) & Uncertainty (1$\sigma$) & Bias/$\sigma$ & Bias/$\sigma$ & Bias/$\sigma$\\
\hline
$\Omega_{\rm m}$& 0.0083 & 0.0093 & 0.016 & -0.035 & -0.032 & -0.0056\\
$\Omega_{\rm b}$& 0.0089 & 0.0094 & 0.013 & 0.079 & 0.068 & 0.022\\
$h$& 0.022 & 0.027 & 0.058 & -0.053 & -0.0044 & -0.00077\\
$n_{\rm s}$& 0.015 & 0.017 & 0.041 & 0.28 & 0.24 & 0.036\\
$\sigma_8$& 0.031 & 0.032 & 0.047 & 0.083 & 0.082 & 0.017\\
$w_0$& 0.17 & 0.19 & 0.33 & 0.059 & 0.046 & 0.024\\
$w_a$& 0.59 & 0.68 & 1.18 & -0.081 & -0.064 & -0.021\\
\hline\hline
\end{tabular}
\end{table}
\twocolumngrid
\subsection{\label{subsec:appkcut}$k$-cut Cosmic Shear and Reduced Shear for Stage IV Surveys}

We calculated the cosmological parameter constraints, and the biases resulting from neglecting the reduced shear approximation, for a range of $k$-cut values. The changing constraints are shown in Figure \ref{fig:sig_kcut}, whilst the biases are shown in Figure \ref{fig:bias_kcut}. As expected, taking lower $k$-cuts results in weaker constraints. In general, biases reduce as a lower $k$-cut is taken. The behaviour of the bias in $\Omega_{\rm b}$ is non-trivial, due to the complex way in which this parameter interacts with the non-linear component of the matter power spectrum. A bias is considered significant if its magnitude is greater than 0.25$\sigma$, as beyond this the confidence contours of the biased and unbiased parameter estimates overlap by less than 90$\%$ \cite{Mas13}. The maximum $k$-cut required in order to ensure that no parameter biases are significant is 5.37 $h$Mpc$^{-1}$. Table \ref{tab:unctab} shows the biases and constraints at this $k$-cut, as well as the biases and constraints when no $k$-cut is taken for both the `optimistic' ($\ell_{\rm max}$=5000), and `pessimistic' ($\ell_{\rm max}$=1500) scenarios for a \emph{Euclid}-like survey. From this, we see that the optimum $k$-cut increases the size of all of the parameter constraints by less than 10$\%$. This is a marked improvement over the `pessimistic' case in which all but two of the parameters have their constraints increased by more than 10$\%$ compared to the `optimistic' case. 

These findings support the idea that $k$-cut cosmic shear can be successfully used to access smaller angular scales for upcoming Stage IV weak lensing surveys. It has already been shown that this technique can bypass the need to model baryonic physics \cite{kcutpap}, while allowing access to small physical scales. Now, these results indicate that $k$-cut cosmic shear can also address the impact of the reduced shear approximation. While explicit calculation of the reduced shear correction yields the most precise cosmological parameter constraints, it is prohibitively computationally expensive \cite{Deshpap}. The $k$-cut approaches bypasses this cost while only marginally weakening the constraints.

We note that if the photometric redshifts are systematically mis-calibrated, the BNT transform we compute would be inaccurate. In fact, given that the lensing kernels have some width, using the peak of the kernel as a representative comoving distance value for the $k$-cut is already technically inaccurate. Despite this, the $k$-cut technique proves successful \cite{kcutpap}. Given that we would expect any biases in the photometric redshifts to be narrower than the width of the kernel, we do not anticipate that these biases would significantly affect the validity of the $k$-cut method. In addition, if there is no mis-calibration, the BNT transformed cross-spectra should be small, and dominated by shot-noise, which is well known and cosmology-independent. If there is significant photometric redshift calibration bias, these cross-spectra will no longer be small. Accordingly, the BNT transform can also serve as a null-test for mis-calibration.

Furthermore, another consideration is our choice of IA model. The NLA model used here can be overly restrictive, and artificially improve constraining power. This could lead to an overestimate of the biases, and accordingly the determination of a lower than needed $k$-cut. However, in any case the limiting $k$-cut value will be that necessitated by baryonic physics.

\subsection{\label{subsec:kinematicappkcut}$k$-cut Cosmic Shear and Reduced Shear for Kinematic Weak Lensing Surveys}

The predicted cosmological parameter constraints for a hypothetical kinematic lensing survey which includes $\ell$-modes up to 5000, together with the expected biases in those constraints from neglecting the reduced shear approximation, are stated in Table \ref{tab:kintabunc}. From this we see that the reduced shear correction is also necessary for potential future kinematic lensing surveys, as the bias in $n_{\rm s}$ is significant. This is due to the fact that constraint on $n_{\rm s}$ is improved, compared to the standard Stage IV case. The spectral index is most sensitive to high-$\ell$ modes \cite{nsexpl}, and this is where the hypothetical kinematic survey performs better than the standard survey. The kinematic survey has a higher signal-to-noise ratio at high-$\ell$, and a lower signal-to-noise ratio at low-$\ell$, as the shot-noise is low by construction, and because it covers a smaller area than the Stage IV survey which means sample variance is relatively more important.

For such a survey, we find that the maximum $k$-cut required for the biases from the reduced shear correction to no longer be significant is 5.82 $h$Mpc$^{-1}$. This is higher than the value in the Stage IV survey case, because the kinematic survey is less deep in redshift. Consequently, the same $\ell$-mode corresponds to a higher $k$-mode for the kinematic survey than in the Stage IV experiment case. Since the the reduced shear correction is only non-negligible at the highest $\ell$-modes, this is where a cut will alleviate biases, and shallower surveys can include higher $k$-modes before reaching this regime. Table \ref{tab:kintabunc} shows the predicted parameter constraints and reduced shear biases at this $k$-cut. For comparison, the constraints and biases for the pessimistic case of the kinematic survey, where only $\ell$-modes up to 1500 are probed, are also shown here. As with the Stage IV cosmic shear survey, the $k$-cut technique degrades the predicted cosmological constraints for a kinematic lensing survey less than the exclusion of $\ell$-modes above 1500. With the $k$-cut, the largest increase is on the constraint on $h$, which increases by 27$\%$. In comparison, in the pessimistic case, the lowest increase in constraints is of 44$\%$, for $\Omega_{\rm b}$.

\section{\label{sec:conclusions}Conclusions}

In this paper, we have examined the validity of the reduced shear approximation when applying $k$-cut cosmic shear to Stage IV cosmic shear experiments, and a hypothetical kinematic lensing survey. We first compared the reduced shear correction calculated using two different models for the matter bispectrum: the fitting formulae of \cite{ScocCouch}, and the \texttt{BiHalofit} model \cite{bihalofit}. Despite the differences between the two approaches, we found that their resulting reduced shear corrections were not significantly different, and that accordingly the reduced shear correction was robust to the choice of bispectrum model.

The $k$-cut cosmic shear technique is used to remove sensitivity to baryonic physics, while allowing access to small physical scales. We examined whether it would also affect the impact of the reduced shear approximation. A variety of $k$-cuts were applied to the BNT transformed theoretical shear power spectra and reduced shear corrections for the `optimistic' case of a \emph{Euclid}-like survey. This scenario assumes $\ell$-modes up to 5000 are probed. We demonstrated that, in this case, $k$-cut cosmic shear preferentially removes scales sensitive to the reduced shear approximation, reducing it's importance. This technique makes this `optimistic' scenario more achievable, while bypassing the significant computational expense posed by having to explicitly calculate the reduced shear correction. The disadvantage is that the inferred cosmological parameter constraints are weakened. However, with $k$-cut cosmic shear applied to the `optimistic' case, the parameters constraints are weakened significantly less than those found in the `pessimistic' case for such a survey; where only $\ell$-modes up to 1500 are included. We also repeated this analysis for a theoretical kinematic lensing survey; finding similarly that the $k$-cut technique reduced sensitivity to the reduced shear approximation.

\begin{acknowledgments}
The authors would like to thank Eric Huff for providing the theoretical galaxy distribution for a hypothetical TF-Stage III kinematic lensing survey. ACD wishes to acknowledge the support of the Royal Society. PLT acknowledges support for this work from a NASA Postdoctoral Program Fellowship. Part of the research was carried out at the Jet Propulsion Laboratory, California Institute of Technology, under a contract with the National Aeronautics and Space Administration.
\end{acknowledgments}

\bibliography{apssamp}

\end{document}